\documentclass[twocolumn,showpacs,preprintnumbers,amsmath,amssymb]{revtex4}
\usepackage{graphicx}
\begin{document}
\def \be{\begin{equation}}
\def \ee{\end{equation}}
\def \bea{\begin{eqnarray}}
\def \eea{\end{eqnarray}}
\def\beps{\mbox{\boldmath $\mathbf{\epsilon}$ }    }
\def\bxi{\mbox{\boldmath $\mathbf{\xi}$ }    }
\def\bbeta{\mbox{\boldmath $\mathbf{\eta}$ }    }
\def\bx{{\bf x}}
\def\cN{{\cal N}}
\def\by{{\bf y}}
\def\bs{{\bf s}}
\def\br{{\bf r}}
\def\bE{{\bf E}}
\def\bA{{\bf A}}
\def\bD{{\bf D}}
\def\bq{{\bf q}}
\def\bd{{\bf d}}
\def\bn{{\bf n}}
\def\bl{{\bf l}}
\def\bj{{\bf j}}
\def\bz{{\bf z}}
\def\cO{{\cal O}}
\def\half{{1\over 2}}
\title{Steady States of a Microwave Irradiated Quantum Hall Gas}
\author{Assa Auerbach$^1$, Ilya Finkler$^2$, Bertrand I. Halperin$^2$  and Amir Yacoby$^3$,}
\address{$^1$Physics Department, Technion, Haifa 32000, Israel.\\
$^2$Physics Laboratories, Harvard University, Cambridge, MA 02138,USA.\\
$^3$Department of Physics, Weizmann Institute of Science, Rehovot 76100, Israel.}
\date{\today}
\begin{abstract}
We consider effects of a long-wavelength  disorder potential on the Zero Conductance State (ZCS) of the 
microwave-irradiated 2D electron  gas.   Assuming a uniform Hall conductivity, we construct a Lyapunov 
functional and derive stability conditions on the domain structure of the photo-generated fields. 
We solve the resulting equations for a general one-dimensional and certain
two-dimensional disorder potentials, and find non-zero conductances,
photo-voltages, and circulating dissipative currents.
In contrast, weak white noise disorder does not destroy the ZCS, but induces  mesoscopic current fluctuations.
\end{abstract}

 \maketitle
\narrowtext
The observation of giant magnetoresistance oscillations in a
microwave-irradiated two-dimensional electron gas (2DEG) \cite{ZRS-exp},
has spurred intensive theoretical activity.
Two distinct microscopic mechanisms for conductivity corrections have been proposed:
(i) The displacement photocurrent (DP)\cite{ZRS-DP},  which is caused by
photo-excitation of electrons into displaced guiding centers and (ii)
the distribution function  (DF) mechanism,  which involves redistribution of
intra-Landau level population for large inelastic lifetimes\cite{ZRS-DF}.

Andreev,   Aleiner and Millis\cite{ZRS-macro}  have noted that irrespective of microscopic details,
once the radiation is strong enough to render the {\em local} conductivity negative,
the system as a whole will break into domains of photogenerated fields and spontaneous Hall
currents. In their proposed domain phase, motion of domain walls
can accommodate the external voltage, resulting in  a
Zero Conductance State (ZCS) in the Corbino geometry, or  a
Zero Resistance State for the Hall bar geometry, in apparent agreement with
experimental reports\cite{ZRS-exp}.
However, one may ask, what should be the effects of long-wavelength (relative to the cyclotron radius) disorder, 
which is either naturally present or deliberately introduced? What is the nature of the coupling between a disorder potential and the photogenerated fields 
\cite{ZRS-disorder},
and could  the disorder pin domain walls?  Such pinning would
affect the macroscopic transport and could destroy the ZCS.

\begin{figure}[htb]
\begin{center}
\includegraphics[height=5cm,width=8cm,angle=0]{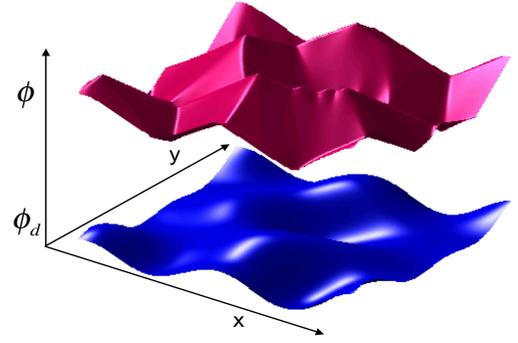}
\caption{Domain phase in the presence
of disorder: the photogenerated potential $\phi(x,y)$ corresponding to a
generic long-wavelength disorder potential $\phi_d$. The steady state is determined
by numerically minimizing the  Lyapunov functional (see text). The domain walls (potential edges) are pinned
by $\phi_d$, yielding a finite conductance.
} \label{Fig1}
\end{center}
\end{figure}

In this Letter  we incorporate  a long-wavelength  disorder potential   $\phi_d(\bx)$ 
into the non-linear magneto-transport equations. We explore its effects on the domain structure and macroscopic transport coefficients. 
For the case of a constant Hall conductivity, we construct
a {\em Lyapunov functional} \cite{Lyapunov} which greatly simplifies the determination of
the stable steady states and their  conductance. We use it to derive general stability conditions on domain walls
in the strong radiation regime.  We also show that  weak 'white-noise' disorder is an {\em irrelevant} perturbation, which does 
not destroy the ZCS. It does introduce, however, mesoscopic non linear current  fluctuations. 
We find solutions for the following disorder potentials: (i) The {\em  general one-dimensional} 
 potential, where domain walls are pinned to the potential extrema, which results in a non zero conductance and  photo-voltage.
(ii) The simple ``egg-carton'' potential solved variationally,  and (iii)  a generic  non-separable potential depicted in Fig.~\ref{Fig1}, which is solved numerically.
(ii) and (iii)  exhibit  two-dimensional domain wall pinning and {\em frustration} effects, which result in circulating dissipative currents.

{\em Non-Linear Magneto-Transport.}
In the presence of an external microwave field,  we  use a local relation between the dc current $\bj(\br)$ and the local
electrostatic field $\bE(\br)$:
\be \bj=\bj^{d}(\bE,\br)+ \sigma^H \hat{\bz}\times \bE \ .
\label{curr}\ee
Here we assume at the outset that the Hall conductivity $\sigma^H$ is a constant,
independent of $\br$ and $\bE$, which  leads to considerable simplifications.
The  dissipative conductivity,  $\sigma^d_{\alpha \beta}(\bE) \equiv \partial j^d_\alpha / \partial E_\beta$,
satisfies
\be
\sigma^d_{\alpha \beta}(\bE,\br) = \sigma^d_{ \beta \alpha}(\bE,\br).\label{jdiss}
\ee

The vector function $\bj^d$, in general, will depend explicitly on the position $\br$, due, e.g., to
inhomogeneities  in the 2DEG, and its direction may not be perfectly aligned with $\bE$.
Eq. (\ref{curr}) is supplemented by the continuity equation,   $\nabla \cdot \bj= -\dot{\rho}$,
where $\rho$ is the charge density.
We emphasize that equation (\ref{curr}) contains all microscopic interactions at lengthscales shorter than the cyclotron radius $l_{c}$, 
which serves as an ultra-violet cut-off, of order 1$\mu$m.

Writing $\bE \equiv - \nabla \phi$,
we may relate changes in the electrostatic potential $\phi$
to changes in $\rho$ through the inverse capacitance matrix $W$:
\be \delta \phi(\br)=\int d^2r'  W(\br,\br') \delta \rho(\br').
\label{ES}
\ee
If a time-independent steady state is reached, then we have simply $\nabla \cdot \bj=0$,
and the precise form of $W$ is unimportant.

In a Corbino geometry, one specifies the potential on the inner and outer boundaries of the sample,
and one looks for a solution for $\phi(\br)$ consistent with these boundary conditions.
Since we asssume $\sigma^H$ to be a constant,  the Hall current cannot contribute to
$\nabla \cdot \bj$ in the interior of the sample, so it does not appear in Kirchoff's equations.
Consequently, the solution for $\phi(\br)$ is independent of $\sigma^H$ and we may, for simplicity set $\sigma^H=0$.
To recover the Hall current, one inserts the
solution $\bE$ into
the second term in (\ref{curr}).

Condition  (\ref{jdiss}) on $\bj^d$ allows us  to  define a scalar {\em Lyapunov functional} as
\be
G[\phi]= \int d^2r g(\bE),~~~g= \int_0^{\bE(\br)} d\bE' \cdot \bj^d(\bE').
\label{G-def}
\ee
A variation of (\ref{G-def}) is given by
\be
\delta G =\int d^2r  \nabla \cdot \bj^d~\delta \phi
-\int_{bound}ds\hat{\bn} \cdot \bj^d \delta\phi . \label{dG} \ee

The second integral vanishes on equipotential boundaries, or in the
absence of external currents.  The extrema of $G$ are found to be steady states, with  $\nabla \cdot \bj=0$.
Using the positivity of the inverse capacitance matrix $W$,
one may show  that $G[\phi(t)]$
is indeed a Lyapunov functional, i.e. a non-increasing function of
time, so that its minima are stable steady states. In general, $G$ may have multiple minima.
Any initial choice of $\phi(\br)$  will relax to some local minimum of $G$,
but not necessarily the ``ground state'' with lowest $G$.  Nevertheless, we expect that in the presence of noise,
the system might tend to escape from high-lying minima and wind up in a state with 
$G$ close to the absolute minimum.

Using the boundary term in (\ref{dG}), the current across a Corbino
sample is equal to the first derivative of $G$ with respect to the
potential difference $V$ between two edges, and the differential conductance
is given by the stiffness, or the second derivative:
\be {d I\over dV} = {d^2 G\over d V^2} .
\label{conduct} \ee

{\em The Domain Phase.}
We now consider a homogeneous system, in the regime of
strong microwave radiation at frequencies slightly larger than
the cyclotron harmonics $\omega>m\omega_c, m=1,2,\ldots$ i.e. positive detuning.
Both DP and DF mechanisms produce a regime of negative conductivity $\bj(\bE)\cdot\bE<0$, which
implies a minimum of $g(E)$ at a finite field $|\bE|=E_c > 0$, which was estimated\cite{VA}  to be of order $\hbar\omega_c /(e l_c)$.
In order to satisfy equipotential
boundary conditions, and the constraints  $\oint d\bl\cdot\bE=0$, field discontinuities  
and charge density singularities must form.

A second order expansion of $g$  about  $E_c$ reads as
\be
g_0(\bE) = g_0(E_c) + \half  (E-E_c)^2 \sigma_c + \lambda |\nabla\cdot \bE|^2 \ . \label{lyap-clean}
\ee
The clean system of Eq.(\ref{lyap-clean}),  is governed by  a `Mexican hat' Lyapunov density,
with a flat valley along $|\bE|=E_c$, i.e. the steady state local conductivity is `marginally' stable everywhere except inside the domain walls. 
The field-derivative coefficient $\lambda \approx \sigma_c  l_{dw}^2$ implements the ultra-violet cut off,
introducing a domain-wall thickness scale $l_{dw}$ assumed here to be of the order of $l_c$.
Domain walls yield a positive contribution to $G$ of order
$\sigma_c E_c^2  l_{dw}$ per unit length. In the absence of disorder, the system will simply minimize  
total domain walls contribution, subject to aforementioned constraints.

A change in the average field $\langle \bE \rangle$, required if there is a change in the applied voltage $V$, can be accommodated by 
a motion of domain walls, or a  reorientation of the local $\bE$.  The relative corrections to zero conductance vanish as $l_{dw}$  over the sample length.
This  defines the clean ZCS phase described in Ref.~\cite{ZRS-macro}.

{\em Long-Wavelength Disorder.} In an inhomogeneous system, there will be a non-zero electrostatic field, $\bE_d(\br) \equiv  - \nabla \phi_d(\br)$, 
present in the thermal equilibrium state, with no microwave radiation or bias voltage. We may ask how this disorder field will alter the ZCS.

At weak disorder field, $|\bE_d|<<E_c$,   the Lyapunov density near 
 $E\approx E_c$ is  modified to 
\be
g(\bE,\bE_d)= g_0(\bE) -   \sigma_1(E)  ~\bE \cdot \bE_d (\br)+\cO(E_d^2),
\label{Lyap}
\ee
which yields a current density
\be
\bj^d(\br) =  -\sigma_1 \bE_d +
{g_0'-\sigma_1'   \bE\cdot\bE_d \over E}\bE,
\label{Curr}
\ee
where $X'\equiv \partial X/\partial E$, and the
coefficient $\sigma_1$ depends on microscopic mechanisms. 

We wish to elaborate on a physical issue regarding Eq.(\ref{Curr}):
In  the non-irradiated  (dark)  linear response theory, the current 
is driven by the {\em electrochemical} gradient $\beps=\bE-\bE_d$. Similarly, one may expect the photocurrent of the DF  mechanism
to also depend on $\beps$.  In contrast, the 'upstream' photocurrent, pumped by the DP  mechanism, 
involves transitions between single particle states which feel the local {\em  electric} field $\bE(\br)$.   Thus, 
due to both contributions, even if the DP mechanism is relatively weak, $\bE_d$ {\em cannot}
be eliminated from Eq.(\ref{Curr}) by a change of variables  $\bE\to\beps$. By our  microscopic estimate\cite{elsewhere}, for the pure DP mechanism,  
$\sigma_c,\sigma_1(E_c)$  are close to the dark conductivity, and $\sigma_1' << \sigma_c /E_c$.

Local stability requires that   $\sigma^d(\br)$, of Eq.~(\ref{jdiss}), has non-negative eigenvalues.
The lower (transverse)  eigenvalue is given by
\be
\sigma_{-}={g_0'-\sigma_1'   \bE\cdot\bE_d \over E} + \cO(E_d)^2 \ge 0 \ ,
\label{cond-eig}
\ee
so  marginal stability occurs at $E=E_c+ \sigma_1' \bE_d\cdot\bE/\sigma_c$.

In a steady state, the normal current density  (in direction $\hat{\bn}$) is continuous across a domain wall. If
$\bE_1$ and $\bE_2$ are the fields on its two sides, we find by
(\ref{cond-eig}) and(\ref{Lyap})  that
\be
\sigma_{-}(E_1) ~\bE_1\cdot\hat{\bn}=
\sigma_{-}(E_2) ~\bE_2\cdot\hat{\bn} ~+\cO(E_d^2).
\label{MS}
\ee
When $\bE_i \cdot \hat{\bn},i=1,2$ have opposite signs (as they do in the clean limit) (\ref{MS}) can only be satisfied
for $\sigma_{-}(E_1)=\sigma_{-}(E_2)=\cO(E_d)^2$.  {\em This restricts the fields at the domain wall to be
at their respective marginally stable values}.
As a result, the current density (\ref{Curr}) at the domain wall reduces to
\be
\bj = -\sigma_1(E_c)\bE_d +\cO(E_d)^2
\label{Curr-DW}
\ee

By  Eq.(\ref{Curr-DW}),  current conservation and Gauss' theorem,  and we obtain a global condition on any closed domain walls,
\be  
 0=\oint\bn\cdot \bj =  -2 \pi \sigma
Q^{2D}+\cO(E_d^2)
 , \label{neutrality}\ee
where $Q^{2D}$ is the integral of the ``2D disorder-charge density,'' $\nabla \cdot \bE_d / 2 \pi $,
over the area enclosed by the loop.

Finally, we note that {\em{generically}}, 
the differential conductance of a sample in the Corbino geometry can be obtained by solving for the 
conductance of a linear system with local conductivity given by $\sigma^d(\bE(\br))$, in series with
 resistive elements along the domain walls, which arise from movement of the domain walls in response 
 to a variation in the applied bias $V$. (There could also be discontinuities in the current at discrete values of $V$, 
 if the system jumps discontinuously from one local minimum of $G$ to another.) We shall see that for weak long-wavelength disorder, 
  the scale of the macroscopic conductance is set by the domain-wall contribution.

{\em White-Noise Disorder}.
In the ZCS, we now  show that a  weak  `white-noise' disorder potential, with a correlation length $\xi_d$ (of the order $l_{dw}\sim l_c$), 
and root-mean-square value ${\bar\phi_d}  \ll E_c \xi_d$    
is an {\em irrelevant} perturbation which does not introduce new domain walls or destroy the ZCS.
This is shown by using an Imry-Ma comparison\cite{ImryMa} of surface to bulk contributions to the Lyapunov functional.
By (\ref{Lyap}), for a square domain of area  $L_d^2$, the negative contribution of
aligning $\bE$ with the averaged
disorder field $\bE_d$, scales as $-\sigma_1 {\bar\phi_d} E_c \sqrt{L_d \xi_d} $.
However the linear cost of  its domain walls grows as  $+\sigma_c E_c L_d l_{dw}$.  
Therefore weak disorder cannot necessarily break the system up into smaller domains.

The disorder, however, will
produce current fluctuations across pre-existing domain walls, needed
to satisfy boundary conditions on the sample. By Eq. (\ref{Curr-DW}),
the current density integrates across the domain wall to yield a random number of order
\be
\delta I  = \pm \sigma_1 {\delta V {\bar\phi_d}\over E_c \xi_d^{5/2}} L_d^{1/2}
\ee
While the conductance will  average out to zero at voltages $V \ge  E_c \xi_d$ or if multiple domains
are in series. The random currents and conductance fluctuations should be observable in small samples, or as harmonic noise generation
for oscillatory bias voltage.

\begin{figure}[htb]
\begin{center}
\includegraphics[height=3.6cm,width=9.1cm,angle=0]{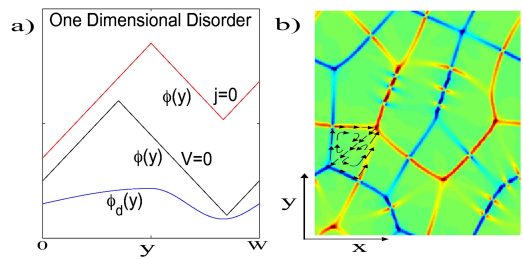}
\caption{a) Photogenerated potential solutions  $\phi$, for one dimensional disorder $\phi_d(y)$, with zero current and  zero voltage boundary conditions. 
b) Map of $\nabla \cdot \bE$ (red-positive, blue-negative) for solution
$\phi (x,y)$ of Fig. 1, showing the domain walls. Circulating dissipative
currents are illustrated in one domain.}\label{Fig2}
\end{center}
\end{figure}  

{\em One-Dimensional Disorder.}
In contrast to weak white-noise disorder, potentials with  long-range correlations can be 
{\em relevant} perturbations. Consider the case of a general  one-dimensional disorder (see Fig.\ref{Fig2}a), 
where $\phi_d(y)$ is independent of the $x$-coordinate. 
At wavelengths larger
than $l_{dw}$ , the Lyapunov functional  is minimized  if the system breaks up into parallel
domains, so that $\bE$ is everywhere aligned with $\bE_d$, and $\bj^d = 0$.
These conditions determine $\bE(y)$  via (\ref{Curr}), and the boundary conditions for a rectangular Corbino geometry 
(periodic boundary conditions at $x=0$ and $x=L$).
 The voltage difference $V$ between the leads at $y=0$ and $y=W$ , satisfies
\be V (I)=  \int_0^W dy (E(y)-E_d(y))  \  .  \label{1Dvoltage}
\ee
At strong  radiation intensity and zero current, (\ref{Curr-DW}) and transverse stability imposes that 
domain walls form precisely at maxima and minima of $\phi_d$, given by the   $y_i ,i=1,\ldots,N$
To lowest order in
$\phi_d$,   there will be a non-zero photovoltage which only depends on these positions:
\be V(0) =  E_c \left( (-1)^N W  -2 \sum_{i=1}^N (-1)^i y_i  \right)  +\cO(E_d),
\label{PV}
\ee
 where  $i=1,3\ldots$ are maxima. The differential conductivity $\sigma_{yy}$ is given, to lowest order in $\phi_d$,  by
\be {1\over \sigma_{yy} } ={2 E_c\over \sigma_1 W}  \sum_i {1\over |\phi_d''(y_i)|}
\equiv {2E_c \over \sigma_1 \tilde{E}_d  } \ ,  \label{resis-1d}\ee
If $P(E_d)$ is the probability distribution for $E_d$ at a random point, it can be shown that  
 $\tilde{E}_d^{-1} = P(0)$.
If $E_d$ is taken from a Gaussian distribution, 
then $\tilde{E_d} = (2
\pi)^{1/2} E_d^{\rm{rms}} $, where $ E_d^{\rm{rms}} $ is the
root-mean-square value of $E_d$.
If $\phi_d$ is a single sine wave, then 
$\tilde{E_d} = 2^{1/2} \pi E_d^{\rm{rms}}$. The transverse differential conductivity 
$\sigma_{xx}$ can also be calculated, using (\ref{cond-eig}), and is given, to first order in $\phi_d$ by 
$\sigma_{xx} = \sigma_1  \langle |\bE_d|\rangle / E_c $.
 For a Gaussian distribution, one has $\sigma_{xx}= ( 2 / \pi) \sigma_{yy}$, while for a single sine wave, 
$\sigma_{xx}= ( 4 / \pi^2) \sigma_{yy}$.  

{\em Two-Dimensional Potentials}
The simplest 2D choice for $\phi_d$  is the separable `egg-carton' potential, given by
\be \phi_d = {E^0_d \over \sqrt{2}} \left( \cos ( x ) +   \cos (y)\right) \  . 
\label{EC}\ee
We  construct a zeroth order trial solution for zero bias current $\bj^d_0=0$  by placing domain walls on
the lines $x=n\pi $ and $y=m\pi$, for integer $n$ and $m$. 
This yields constant electric fields in each square domain, of the form:
\be \bE_0 = {E_c \over \sqrt{2}} (\pm\hat{x}   \pm\hat{y}  ) \ . 
\label{trial}
\ee
$\bE_0$ is a gradient of a continuous potential, and satisfies charge neutrality (\ref{neutrality}). 
Upon application of an external voltage in an arbitrary direction,  
domain walls will move, as in the one-dimensional case, and also tilt with $\bE$ into a
herringbone pattern. A variational calculation, assuming that within each domain 
$\bE$ is constant, finds, to first order in
$E_d$, that $\sigma_{xx} = 0.83 \sigma_1 E^0_d / E_c $.  Numerical 
calculations confirm that the variational solution is at least close to the 
exact answer.

We have calculated analytically the first order (in $E_d$) corrections to
(\ref{trial}), for zero external voltage, by integrating
Kirchoff's laws\cite{elsewhere}.  Away from the domain walls, we find $E >E_c $, which corresponds to
circulating dissipative currents,  which match onto
the tangential currents at domain walls, given by Eq. \ref{Curr-DW}.
In Fig.~\ref{Fig1},  a generic  two dimensional example is displayed. $\phi_d$ contains 20 Fourier components
chosen from Gaussian distributions with  $<|\phi_d({\bf{k}}|^2> $ independent of ${\bf{k}}$.  
The  potential $\phi$ is found by numerically minimizing $G$.   Fig. ~\ref{Fig2}b   plots the 2D charge density where domain walls
appear as line singularities.  
In both two dimensional examples, (\ref{EC}) and  Fig.\ref{Fig1},  $G$ is  {\em frustrated} from perfect  alignment of $\bE$ and 
$\bE_d(\br)$ by the conditions $|\bE|\ge E_c$ and $ \nabla\times \bE=0$.  This frustration underlies the circulating dissipative currents
which are illustrated in one of the domains in Fig.~\ref{Fig2}b.

In summary, we have introduced long wavelength disorder into the transport theory of the microwave
irradiated quantum Hall gas, using the Lyapunov functional as an organizing principle for the stability of  steady states.
We showed  that weak white noise disorder is  irrelevant for
the stability of the ZCS although it produces mesoscopic current fluctuations.
For a strong and long range potentials, the ZCS state breaks  up into  domains, 
which will generally result in a finite conductance and a photovoltaic effect. A microscopic theory necessary for
the steady states dependence on microwave power and detuning frequency is deferred to a forthcoming publication\cite{elsewhere}.
We have not considered effects of conductivity anisotropy and
variations in Hall coefficient. The latter will not affect the domain pattern or
the longitudinal conductivity for one-dimensional disorder, but  might  have large effects 
and be experimentally relevant in other geometries.

{\em Acknowledgment}. We thank E. Meron,  A. Stern, R. L. Willett  and Y. Yacoby for helpful discussions.
AA and AY are grateful for the hospitality of Harvard University,  Aspen Center for Physics and Kavli Institute for Theoretical Physics.
Work was supported in part by the Harvard Center for Imaging and Mesoscale Structures,  NSF grant DMR02-33773, 
the US-Israel Binational Science Foundation, and the Minerva Foundation.


\begin{thebibliography}{99}
\bibitem{ZRS-exp} R. G. Mani et.al. Nature, 420, 646 (2002);
M. A. Zudov et. al. Phys. Rev. Lett. 90, 046807 (2003);
C. L. Yang et. al. Phys. Rev. Lett. 91, 096803 (2003);
R. L. Willett, et. al. Bull. Am. Phys. Soc. 48, 459 (2003).


\bibitem{ZRS-DP}
A. Durst, S. Sachdev, N. Read, and S. M. Girvin, Phys. Rev. Lett. 91, 086803
(2003); V. I. Rig, Fiz. Tverd. Tela 11 , 2577 (1969) [Sov. Phys. Solid State 11 , 2078 (1970)];
P.W. Anderson and W.F. Brinkman, cond-mat/0302129;
J. Shi and X.C. Xie, Phys. Rev. Lett. 91, 086801 (2003).
\bibitem{ZRS-DF} I. A. Dmitriev, A. D. Mirlin and D. G. Polyakov, Phys. Rev. Lett. 91, 226802 (2003);
I. A. Dmitriev, M.G.Vavilov, I. L. Aleiner, A .D. Mirlin and D. G. Polyakov, preprints cond-mat/0310668 and 0409590.
\bibitem{ZRS-macro} A.V. Andreev, I.L. Aleiner, and A.J. Millis, Phys. Rev.
Lett. 91, 056803 (2003);


\bibitem{ZRS-disorder} Effects of a {\em short wavelength}  periodic potential have been
recently addressed by
by J. Dietel, L. Glazman, F. Hekking, F. von Oppen,
cond-mat/0407298.

\bibitem{Lyapunov} M.C. Cross and P.C. Hohenberg, Rev. Mod. Phys. 65, 851 (1993). The dynamical phase transition
in J. Alicea, L. Balents, M.P.A. Fisher, A. Paramekanti, L. Radzihovsky, cond-mat/0408661, is governed by a Lyapunov functional.

\bibitem{VA} M.G. Vavilov and I.L. Aleiner, Phys. Rev. B 69, 035303 (2004).

\bibitem{elsewhere} A. Auerbach,  I. G. Finkler, B.I. Halperin, A. Yacoby, in preparation.

\bibitem{ImryMa} Y. Imry and S.-k. Ma, Phys. Rev. Lett. 35, 1399 (1975).


\end{thebibliography}
\end{document}